# Is the luminosity distribution of field galaxies really flat ?

Simon P. Driver

School of Physics, University of New South Wales,

Sydney, NSW 2052, Australia

Steven Phillipps

Astrophysics Group, Department of Physics,

University of Bristol, Tyndall Road, Bristol BS8 1TL, UK

## ABSTRACT

Recent observations of the galaxy population within rich clusters have found a characteristic luminosity distribution described by a flat ($\alpha = -1.0$) Schechter function which exhibits an upturn at faint absolute magnitudes ($M_B \approx -18$). Here we discuss whether such a form for the field luminosity distribution is ruled out by local and/or faint magnitude limited redshift surveys (MLRS).

Our conclusions are that existing redshift surveys provide little constraints on the volume-density distribution of field galaxies faintwards of $M_B = -18$. The local MLRS suffer from poor statistics over inhomogeneous volumes, while the faint MLRS are ambiguous because of the unknown nature of the "faint blue excess" and the "normalization" problem.

Adopting a functional form similar to that seen in rich clusters we find that the maximum allowable faint end slope, based on the Mt Stromlo-APM redshift survey, is $\alpha \approx -1.8$ faintwards of $M_B = -18.0$ ($H_o = 50$ kms$^{-1}$ Mpc$^{-3}$).

*Subject headings:* galaxies: distances and redshifts — galaxies: luminosity function

## 1. INTRODUCTION

The typical density of galaxies in a homogeneous volume is a fundamental ingredient in numerical models of the Universe. The accuracy to which we can determine this distribution ultimately defines our ability to understand the distant galaxy population and the changes which may have (and must have) occurred with look-back time. In particular the density of the field dwarf population (here we define dwarfs as *all* galaxies with absolute magnitudes fainter than $M_B = -18 + 5\log h_{50}$) becomes increasingly important as we look to fainter apparent magnitudes (c.f. Kron 1982; Koo & Kron 1992; Driver *et al.* 1994a). With photometric observations now reaching as faint as $b_J \sim 27.5$ (Metcalfe *et al.* 1995) and the majority of faint galaxy models invoking evolution of the low luminosity population (e.g. Phillipps & Driver 1995 and references therein) understanding the underlying local density of dwarf galaxies becomes vital. What do we know about the size of this population and how well is it currently constrained ?

In §2 we summarize the difficulties of determining the faint end slope of the field population from magnitude limited redshift surveys (MLRS), in §3 we collate the recent observations of cluster environments, which typically show an upturn at the dwarf/giant boundary, and in §4 & §5 we determine the upper limit to the local density of dwarf galaxies from the existing bright and faint redshift surveys. Finally in §6 we present our conclusions.

## 2. THE FIELD LUMINOSITY DISTRIBUTION

The currently favored description for the luminosity distribution of field galaxies is that of the Schechter luminosity function (Schechter 1976) which is described by three fundamental numbers. These are (i) the intrinsic luminosity of the $L_*$ galaxy (at the "knee" of the function), (ii) $\phi_*$ the scaled number density of $L_*$ galaxies, and (iii) $\alpha$ the faint slope parameter, as shown below, (see Schechter 1976; Felten 1985; Binggelli, Sandage & Tammann 1988).

$$\phi(L)dL = \phi_* \left(\frac{L}{L_*}\right)^\alpha e^{-(\frac{L}{L_*})} d\left(\frac{L}{L_*}\right) \qquad (1)$$

The most straightforward method for determining these parameters is to measure the redshifts for a complete magnitude limited sample of galaxies and apply a suitable fitting algorithm[1] (e.g., Efstathiou, Ellis & Peterson 1988, EEP; de Lapparent *et al.* 1989; Loveday *et al.* 1992 and Marzke *et al.* 1994a, b). The first three of these surveys, find comparable Schechter parameters of, $\phi_* \approx 0.002$ Mpc$^{-3}$, $M_*^B \approx -21.0$ and $\alpha \approx -1.05 \pm 0.10$ ($H_o = 50 kms^{-1}Mpc^{-1}$) and these have been the typical values adopted in many numerical models (e.g. Rocca-Volmerange & Guiderdoni 1990; Yoshii 1993; Glazebrook *et al.* 1994 etc). Marzke *et al.* also find $\alpha \approx -1$ over the range $-20 < M_B < -18$ but find both a fainter normalization point of $M_* = -20.3$ and an excess of low luminosity systems in the range $-17.5 < M_B < -14.5$. However, the exact size of this low luminosity excess is unclear, due to significant uncertainty in the Zwicky magnitudes, particularly at the faint end. Together these surveys argue consistently for a flat luminosity distribution over the range $(-20 < M_B < -18)$ which can be well described by a Schechter function with $\alpha \approx -1.0$. But what form does the distribution take at fainter magnitudes ?

There are two related problems in determining the faint end slope from a local MLRS: small number statistics and inhomogeneities. In such a survey, the more luminous galaxies are seen over a larger volume and therefore in greater numbers and any inhomogeneities are averaged out. Conversely for the less luminous galaxies the observed volume is smaller and therefore more prone to the vagaries of small number statistics and inhomogeneities. The direct consequence is that whilst the bright end of the field luminosity distribution can be determined relatively easily, the faint end may have a high degree of uncertainty (which depends critically on the completeness of the sample and the homogeneity of the volume over which the fainter galaxies are observed).

---

[1]Various methods have been devised to verify that the Schechter function provides a representative fit to the data, such as the Stepwise Maximum Likelihood method of EEP, for example, which derives the volume-density distribution of galaxies without assuming a specific functional form.

One way to address the problem is to measure the luminosity distribution for individual clusters, as essentially the field luminosity distribution is an average over many clusters and groups. Measuring the luminosity distribution for individual clusters is substantially easier than for the field as it requires no redshifts and simply involves subtracting the background $\frac{dlogN}{dm}$ distribution from that of a sight-line through the cluster (see Driver *et al.* 1994b for more details).

Recent measurements of the luminosity distribution seen in *rich* clusters have revealed flat distributions over bright magnitudes which apparently turn-up at some intermediate magnitude (e.g. Metcalfe, Godwin & Peach 1994; Godwin, Metcalfe & Peach 1983; DePropis *et al.* 1995; Driver *et al.* 1994b and §3). This discontinuity is generally seen to occur at the point at which dwarf galaxies dominate over the giants (c.f. Sandage, Binggelli & Tammann 1985) and elusively just beyond the limit to which the existing local field MLRSs reach (c.f. §4).

## 3. The Cluster LF

Figure 1 shows the observed luminosity distribution for three clusters: A963 (Driver *et al.* 1994b), A2554 (Smith, Phillipps & Driver 1995), and Coma (Godwin, Metcalfe & Peach 1983). Also shown in Fig. 1, are: a flat Schechter function, a Schechter function with

$\alpha = -1.4$ and a two component Schechter-like function, defined as follows:

$$L_{max} > L > L_{Dwarf} : \quad \phi(L)dL = \phi_* \left(\frac{L}{L_*}\right)^\alpha e^{-(\frac{L}{L_*})} d\left(\frac{L}{L_*}\right)$$

$$L_{Dwarf} > L > L_{min} : \quad \phi(L)dL = \phi_{Dwarf} \left(\frac{L}{L_{Dwarf}}\right)^{\alpha_{Dwarf}} d\left(\frac{L_{Dwarf}}{L_*}\right)$$

$$\text{where}: \quad \phi_{Dwarf} = \phi_* \left(\frac{L_{Dwarf}}{L_*}\right)^\alpha e^{-(\frac{L_{Dwarf}}{L_*})}$$

This two-component LF is adopted as the simplest extension over the normal Schechter function (c.f. Equation 1) as it includes only two additional parameters: $L_{Dwarf}$ to represent the absolute magnitude where dwarfs first dominate over giants (taken here, in the $R$ band, to be at $M^R_{Dwarf} = -19.5$, for $H_o = 50$ kms$^{-1}$Mpc$^{-1}$); and $\alpha_{Dwarf}$ the faint slope parameter for the dwarf population ($\alpha_{Dwarf} = -1.8$ in Fig. 1). The functions shown on Fig. 1 are optimized by eye to fit the A963 data. Only a small number of clusters have so far been studied to sufficient depth and the majority show this steepening of the faint end slope. at fainter absolute magnitudes. The gradient of this turn-up and the point at which it turns up is seen to vary from cluster to cluster (c.f. Fig. 1) suggesting some scatter and/or environmental dependency. As yet the available data is insufficient to warrant detailed comparison, and measurements of the luminosity distribution for a larger number of clusters is currently underway (Phillipps, Driver & Smith 1996; Smith, Driver & Phillipps 1996; Driver *et al.* 1995, in preparation).

The upturn in Coma has been confirmed by Thomson & Gregory (1993) and more recently by Biviano *et al.* (1995) and similar LFs have been observed in other nearby local groups by Ferguson & Sandage (1991), in Virgo by Impey, Bothun & Malin (1988), in Shapley-8 by Metcalfe, Godwin & Peach 1994 and from limited redshift surveys of other local Abell clusters (A2052, A2107, A2199 & A2666) by De Propis *et al.* (1996). Bernstein *et al.* (1995), however, find their data for the *core region* of Coma is more consistent with a single Schechter function with slope $\alpha = -1.4$ ($-15 \leq M_R \leq -11$). Note that this slope is significantly steeper than that found for the field.

If this trend is common, as so far indicated, then we might expect to see a similar upturn in the field LF which is an averaged distribution over many clusters, groups and voids. To investigate further whether such a form is consistent with existing field data, we shall adopt the functional form shown above and determine the constraints on $\alpha_D$ (the dwarf slope) from the current surveys.

## 4. Constraints on $\alpha_{Dwarf}$ from Local MLRS

A limit to $\alpha_{Dwarf}$ can be determined from the bright or local MLRS and here we concentrate on the recent Mt Stromlo-APM survey of Loveday *et al.* (1992), in which the redshifts of $\sim 1800$ galaxies were obtained ($b_J \leq 17.15$). For such a magnitude limited sample the number of galaxies visible at each intrinsic luminosity is given by,

$$\Upsilon(L)\delta L = \phi(L)V(L)\delta L \qquad (2)$$

where,

$\Upsilon(L)\delta L$ is the total number of galaxies visible in a luminosity interval $\delta L$,

$\phi(L)$ is the function which describes the true space density of galaxies per $\mathrm{Mpc}^3$,

and

$V(L)$ is the function which describes the "visibility" or volume within which a galaxy of intrinsic luminosity $L$ can be seen, *i.e.* $\propto L^{\frac{3}{2}}$ over Euclidean distances

The function $\Upsilon(L)$ therefore represents the actual *observed* space density of galaxies with intrinsic luminosity, $L$, in an apparent magnitude limited sample[2]. The most representative way of expressing $\Upsilon(L)$ is simply to plot the number of galaxies observed in real numbers at each intrinsic magnitude. This is preferable to the more typical representation in which log number per Mpc$^3$, versus intrinsic magnitude, is plotted with the "visibility" term de-convolved (*i.e.* $\log_{10} \phi(L)\delta L$ v $M$ as opposed to $\Upsilon(L)\delta L$ v $M$). In order to illustrate how important the "visibility" term is we shall adopt four model LF's:

**a)** the conventionally quoted field LF with $\alpha = -1.0$, (EEP; Loveday *et al.* 1992; de Lapparent *et al.* 1989),

**b)** a single Schechter function with $\alpha = -1.5$ (*i.e.* the typical mean cluster value c.f Sandage, Binggelli, Tammann 1985; Impey, Bothun & Malin 1988, Irwin *et al.* 1990a; Bernstein *et al.* 1995),

**c)** a two component Schechter-like function with a bright slope of $\alpha = -1.0$ and a faint slope value of $\alpha_{Dwarf} = -1.5$, from $M_{Dwarf} = -18.0$ (*i.e.* equivalent to that seen in the field by Marzke *et al.* 1994b and in A2554, c.f. Fig 1.)

**d)** a second two component Schechter-like function identical to *c)* except with $\alpha = -1.8$, (*i.e.* equivalent to that observed in A963, and Coma, c.f. Fig 1.).

Figure 2 (lower panel) shows the "observed distributions" for the four adopted LF's assuming a sample size of $\sim 1800$ galaxies and an apparent magnitude limit of $m_B = 17.15$ (*i.e.* equivalent to the Mt Stromlo-APM survey). Also shown on Fig. 2 (upper panel) are the same four models represented in the more conventional manner as log number density versus absolute magnitude. Note that while the conventional plots are obviously very different the "observed distributions" are less distinct and differences amount to a small number of galaxies. Only case *b)* is readily distinguishable from the observed data and can therefore be ruled out as a possible fit. Case *a)* represents the proposed fit from Loveday *et al.* for the Mt Stromlo-APM survey, yet the "observed distribution" appears equally well described by models *c)* and *d)*, despite their contrasting *volume-corrected* LFs (c.f. Fig. 2, upper panel).

---

[2]To be precise we should replace $\phi(L)$ with $\phi(L, \Sigma)$ which includes the effects of surface brightness ($\Sigma$) (see Disney & Phillipps 1983). However the selection effects associated with $\Sigma$ shall be ignored here and we define visibility only in terms of limiting luminosity (valid for surveys in which the faintest included galaxy is well above the faintest detectable apparent luminosity).

Increasing the size of the survey can overcome the small number statistics, at the cost of telescope time, however the problem of homogeneity is more difficult. Fitting algorithms have been devised which can determine the volume-density distribution of inhomogeneous samples but typically rely on the assumption that all galaxies are clustered similarly, c.f. EEP. If this assumption is invalid then the magnitude limit of the survey ultimately defines the range of the intrinsic luminosities probed. For example, to determine the field LF to $M_B = -18$, requires inhomogeneities to be on scales significantly less than $\sim 100$ Mpc (or 40 Mpc for $M_B = -16$).

The overall implication from Fig. 2 is that the Mt Stromlo-APM survey accurately defines the bright end of the field LF but the slope of the field LF from $M_B > -18.0$ can only be constrained such that $\alpha_{Dwarf} \leq -1.8$, *i.e.* LF *d)*.

## 5. Constraints on $\alpha_{Dwarf}$ from Faint MLRS

The problem of measuring the faint end slope of the luminosity function from a local MLRS has been previously noted (e.g. Phillipps & Shanks 1987) and one method by which it has been addressed is via fainter MLRS (see Broadhurst, Ellis & Shanks 1988; Colless *et* *al.* 1993; Cowie *et al.* 1991 and references therein). The principle is that as intrinsically luminous galaxies are observed at greater distances their observed density rises progressively more slowly than $L^{3/2}$, due to cosmological effects. This is illustrated in Fig. 3 which shows the "inner workings" for our four adopted models (the models are illustrated as log$N$ v $M$ insets in the top left corner of each plot). The model predictions shown in Fig. 3 take no account of surface-brightness effects, seeing or other phenomena and are simply intended to reflect the broader implications of changing the initialy adopted faint end slope. For each panel the bold line represents the prediction of the total galaxy number-counts based on the adopted LF, k-corrections, a standard flat cosmology and no-evolution (c.f. Driver *et al.* 1994a). Each of the remaining lines represents the contribution to the total counts from a narrow luminosity class (solid lines giants, $M_B < -18$; dashed lines dwarfs, $M_B > -18$). Immediately apparent are the cosmological effects, most notably the k-corrections, which cause the flattening of the individual lines from the Euclidean slope of $\frac{dlogN}{dm} = 0.6$. At progressively fainter magnitudes the intrinsically more luminous (and therefore more distant) galaxies are effected more severely, and so the predicted total number-counts, depend more heavily on the contribution from the lower luminosity classes.

Given these trends a faint galaxy survey is expected to place stronger constraints on the faint end slope of the field LF than a local redshift survey. In reality this is not quite so simple, as faint galaxy samples are also critically dependent on evolutionary processes (Tinsley 1980) and on the local normalization problem:

### 5.1. The Faint Blue Excess Problem

The long-standing problem of the excess number of blue galaxies seen at faint apparent magnitudes (c.f. Koo & Kron 1992) makes any comparison between models and data problematic. For instance at $b_J = 23.5$ the standard no-evolution models under-predict the number-counts by a factor of 2–4 (e.g. Broadhurst, Ellis & Shanks 1988; Driver *et al.* 1994a; Metcalfe *et al.* 1995). Returning to Fig 3. we can indeed see that none of the four *no-evolution* models matches the faintest counts which argues convincingly for some strong evolutionary process. Clearly evolution is at work, but as yet its nature and the fate of this faint blue population is unknown and widely speculated upon (merged, faded, dissipated ?). Recent evidence from morphological number counts (Driver, Windhorst & Griffiths 1995; Driver *et al.* 1995; Glazebrook *et al.* 1995), from Mg II absorbers (Steidel, Dickinson & Persson 1994), from the $\Theta - z$ relationship (Mutz *et al.* 1994); and from the moderately faint MLRS (Lilly *et al.* 1995; Ellis *et al.* 1996) together suggest little evolution in the giant populations (Ellipticals through to mid-type Sb-Spirals). This implies that the faint blue excess population is linked to the low luminosity (dwarf/Irregular) population. Understanding the mode and magnitude of this evolution is therefore essential *before* the faint MLRS can be used to define or constrain the faint end slope of the field luminosity distribution.

In so far as comparisons can be made it is *perhaps* valid to assume that at low redshift such evolutionary processes are likely to be small or negligible (but see also Maddox *et al.* 1990 who suggest strong evolution may be occurring locally). On the basis of such an assumption a comparison between the observed data and the no-evolution model predictions at $z < 0.2$, for instance, is then justifiable.

## 5.2. The Local Normalization Problem

The problem of how to normalize models to the data is becoming perhaps more worrisome of late than the problem of the faint blue excess (see discussions in Shanks 1990 and Driver, Windhorst & Griffiths 1995 for example). The problem is summarized by; the steep number-counts at bright magnitudes seen in the APM survey (Maddox *et al.*

1990) and the higher normalizations found in measures of the field luminosity function at $z = 0.1$–$0.3$ (Colless *et al.* 1994; Lilly *et al.* 1995; Ellis *et al.* 1996) compared to those at z=0.0 (Loveday *et al* 1992 etc). The various solutions proposed are: strong local evolution in the giant population (ruled out by the recent HST observations listed above ?); a large local underdensity-inhomogeneity (of radius $\sim 300 - 500$ Mpc); photometric errors in the local samples (c.f. Metcalfe, Fong & Shanks 1995); and/or surface brightness selection effects (c.f. Ferguson & McGaugh 1995). The significance of the normalization problem is that typically faint galaxy models are "scaled up" by a factor of $\sim 2$ to match the number-counts at $b_J = 18$ (considered a sufficiently large distance to be homogeneous but not so large as for evolution to have taken place.). Two potential difficulties are raised by the practise of re-normalization.

1) All of the proposed ideas to explain the steep counts are both morphology and luminosity dependent, implying that a simple "scaling-up" is naive as it does not allow for a "shape correction".

2) At $b_J = 18$ the number-counts are made up of a variety of galaxies over a range of redshift ($0 < z < 0.15$), a simple re-normalization at a fixed apparent magnitude is unrealistic as the normalization is more likely to be linked to redshift than to apparent magnitude.

Given these uncertainties it is not clear exactly how or at which magnitude models should be normalized to the data, one argument in favor of normalizing at $b_J \sim 18$ comes from the recent HST morphological number-counts. Both the Elliptical (E/S0) and Early-type Spiral (Sabc) number-counts are well fit by the no-evolution models when re-scaled by a factor of 2 (equivalent to $b_J \sim 18$ in the total number-counts plot). That both populations require a similar normalization and that both populations are confirmed not to evolve via Mg II absorption studies and the $\Theta - z$ relationship argues for re-normalization at $b_J \sim 18$.

### 5.3. Comparison of Models to the Faint MLRS

Figure 4 shows the most recent redshift survey by Glazebrook *et al.* (1995) giving the observed redshift distribution in the magnitude range $23 < m_b < 24$ for 80 galaxies. Also shown on Fig. 4 are the four models with alternative normalizations to illustrate the magnitude of the "normalization" problem. Fig. 4a shows the models unnormalized (*i.e.* taking the local measure of the field LF normalization at face value, c.f. Loveday *et al.* 1992). In Fig 4b. the models are normalized so as to match the number-counts at $m_b = 18.0$ as argued above (see also Driver *et al.* 1995; Metcalfe *et al.* 1995a) and Fig 4c. is normalized to match the total number of galaxies in the redshift survey, *i.e.* at $b_J = 23.5$ (as suggested by the referee). The contrast between the three panels show the severe ambiguity raised by the "normalization" problem. That none of the models in panels (a) and (b) match the high-z distribution is a reflection of the faint blue excess problem (and not of interest in this paper which seeks only to make a comparison between the no-evolution models and the data at low redshifts). Overall our primary conclusion is that any attempt to constrain the local field luminosity distribution from faint MLRS is open to ambiguity and overshadowed by these other problems. However normalizing the N(z) distribution at $b_J \sim 23.5$, as in Fig 4c, can be ruled out as the normalization is *not* a free parameter but is fixed by the N(m) distribution. *i.e.* To be consistent the N(z) and N(m) distributions must use the same normalization and if this occurred at $b_J = 23.5$ then the models would severely over-predict the observed N(m) distribution at all magnitudes brighter than $b_J = 23.5$ (c.f. Fig. 3a,b,c & d).

We believe that the most reasonable normalization, as suggested by the recent HST observations, is at $b_J \sim 18$, *i.e.* Fig 4b. Table 1 compares the models and the data from Fig 4b under the assumption that no significant evolution is taking place, out to z = 0.2. From Table 1 and Fig 4b we see that both models b) and d) over-predict the z ¡ 0.1 distribution, recall that model b) is already ruled out by the local MLRS. Model c) appears to match the limited available data the best, followed by Model a) the extrapolation of the bright MLRS. However with only 8 galaxies in the sample it is clear that the faint MLRS also suffer from poor number statistics. Furthermore the interpretation of the faint MLRSs is highly ambiguous due to the "normalization" and "faint blue excess" problems. In so far as an upper limit can be placed on the density of local dwarfs from the faint MLRS we conclude that at present this limit is no more stringent than that placed by the local MLRS, $\alpha_D \leq -1.8$.

## 6. Discussion

We have presented a discussion of whether the characteristic luminosity distribution seen in rich clusters is consistent with the bright and faint magnitude limited redshift surveys (MLRS) of the field. By adopting an optimal functional form to match the observed distribution in A963 we find the following:

- The bright MLRSs constrain the distribution of bright galaxies, and these are well described by a single Schechter function with parameters $\alpha \approx -1.0 \pm 0.1, M_* \sim -21, \phi_* \sim 0.002 Mpc^{-3}$ in the range $-20 < M_B < -18$.

- Any turn-up in the field luminosity function at $M_B = -18$, as is seen in A963 and other clusters (c.f §3), is constrained such that $\alpha_D \leq -1.8$.

- The interpretation of the faint MLRS are plagued by two problems: the "normalization" problem and the "faint blue excess" problem.

- Under the assumption of normalizing the models to the data at $b_J = 18$ and assuming any evolution out to $z = 0.2$ is small, we favor a field LF faint end slop in the range $-1.0 \leq \alpha_D \leq -1.5$ beyond $M_B = -18$.

Overall we conclude that the local density of low luminosity systems in the field is poorly constrained and the recent trends seen in rich cluster environments is fully consistent with the available data for the field. *If* the form of the field LF is similar to that seen in rich clusters it may have strong repercussions on the following topics:

1) Faint galaxy number-counts find an excess of galaxies at faint magnitudes over the standard no-evolution model which could be partially explained by a steep faint end slope (Koo & Kron 1993; Driver *et al.* 1994a; Ferguson & McGaugh 1995; Phillipps & Driver 1995).

2) Recent morphological studies (Griffiths *et al.* 1994; Driver, Windhorst & Griffiths 1995; Driver *et al.* 1995; Glazebrook *et al.* 1995) find that the faint excess galaxy-counts are dominated by galaxies with late-type/Irregular appearance.

3) Large-scale galaxy formation models predict steep field LFs (Efstathiou 1995), the reconciliation of such models to the current local observations could be via a two component Schechter function with a steep faint end.

4) The low amplitudes measured for the two point angular correlation function (Efstathiou *et al.* 1991; Couch *et al.* 1993) implies that clustering in the field was significantly less in the past *and/or* that the variety of galaxies seen in a faint apparent magnitude slice derive from a wide range of redshift (c.f. Brainard, Smail & Mould 1995) as would be expected if the field LF turns-up.

5) The contribution of field galaxies to the total baryon density would be increased if the field LF has been underestimated, however such an increase is likely to be small unless the mass-to-light ratios strongly increase with decreasing luminosity (c.f. Persic & Salucci 1990; Bristow & Phillipps 1994).

– 16 –

We conclude that while the space distribution of bright galaxies is well described by a flat Schechter function to $M_B < -18$ the distribution at fainter luminosities is poorly constrained. The maximum allowable upturn in the field luminosity function is $\alpha_D < -1.8$ beyond $M_B = -18$.

We thank Warrick Couch, Rogier Windhorst and Dave Burstein for useful comments and also Nigel Metcalfe for providing the Coma data. SPD thanks the Australian Research Council and notes that much of this work was completed at the University of Wales College of Cardiff and also Arizona State University. SP thanks the Royal Society for financial support via a University Research Fellowship.

TABLES

Table 1: The observed versus the predicted number of galaxies for the four adopted models normalized at $b_J \sim 18$.

| Redshift Range | Observed Number | Predicted Number | | | |
|---|---|---|---|---|---|
| | | (a) | (b) | (c) | (d) |
| 0.00 – 0.05 | 0 | 0.07 | 1.63 | 0.47 | 1.75 |
| 0.00 – 0.10 | 1 | 0.37 | 4.90 | 1.42 | 3.74 |
| 0.00 – 0.15 | 4 | 1.04 | 9.13 | 2.65 | 5.62 |
| 0.00 – 0.20 | 8 | 2.11 | 13.86 | 4.02 | 7.29 |

Figure Captions

**Figure 1** The observed luminosity distributions for the clusters: A963 (Driver *et al.* 1994) A2554 (Phillipps, Driver & Smith 1995) and Coma (Godwin, Metcalfe & Peach 1983). The two dashed lines show conventional Schechter functions with slopes of $\alpha = -1.0$ and $-1.4$. The solid line shows a two component Schechter-like function (see text) with a flat bright slope and a steep upturn of $\alpha = -1.8$ at $M_B \approx -18$ The lines are optimized to fit A963.

**Figure 2** The observed distribution of galaxies in a local magnitude limited sample (bottom panel) assuming four alternate descriptions of the field LF (shown in the top panel). The four distributions are normalized to give 1800 galaxies in the range $-23.0 < M < -16.5$ to be comparable to the Loveday *et al.* (1992) survey. Case *a)* represents the best Schechter function fit to the Mt Stromlo-APM data. The top panel shows the conventional $\frac{d(log_{10}N)}{d(M)} v M$ plots and clearly models a), c) and d) are indistinguishable despite their contrasting implications for the space density of dwarf galaxies. Note that the shaded region on the bottom panel represents the incompleteness in the original Mt Stromlo-APM sample (of 27 galaxies). Any data point in this region therefore contains fewer galaxies than the samples total incompleteness. The errors are $\sqrt{n}$ where n is the number of galaxies in that magnitude interval.

**Figure 3** The inner workings of a typical faint galaxy model for the four adopted models. The thick lines show the total number-count prediction, the other lines show the contribution from an absolute magnitude bin (solid lines for giants $M_B < -18$ and dashed lines for dwarfs $M_B > -18$). Note that at bright magnitudes the total counts depends critically on the intrinsically luminous galaxies while at faint magnitudes the counts depend more critically on intermediate and faint intrinsic luminosities, dependent on the locally luminosity distribution of galaxies (shown as log$N$ v $M$ insets in the upper left corner of each panel. The data are from: Driver *et al.* 1994a (solid triangles); Tyson 1988 (solid squares); Metcalfe *et al.* 1995 (open hexagons); Metcalfe *et al.* 1991 (open triangles); Jones *et al.* 1991 (open squares); Lilly *et al.* 1991 (stars); Shanks 1990 (crosses).

**Figure 4** The observed redshift distributions for a magnitude interval at $23.0 < m_B < 24.0$ (c.f. Glazebrook *et al.* 1995) is shown along with the predictions of four simple models (see text). The three panels show the same data with the four adopted models normalized at: (upper) locally; (middle) at $b_J = 18.0$ and (lower) at $b_J = 23.5$. The three panels illustrate the scale of the "normalization problem". Adopting the middle panel as the most likely comparison we see that models *a)*, *b)* and *c)* all underpredict the local density of dwarfs (*i.e.* $z < 0.2$).

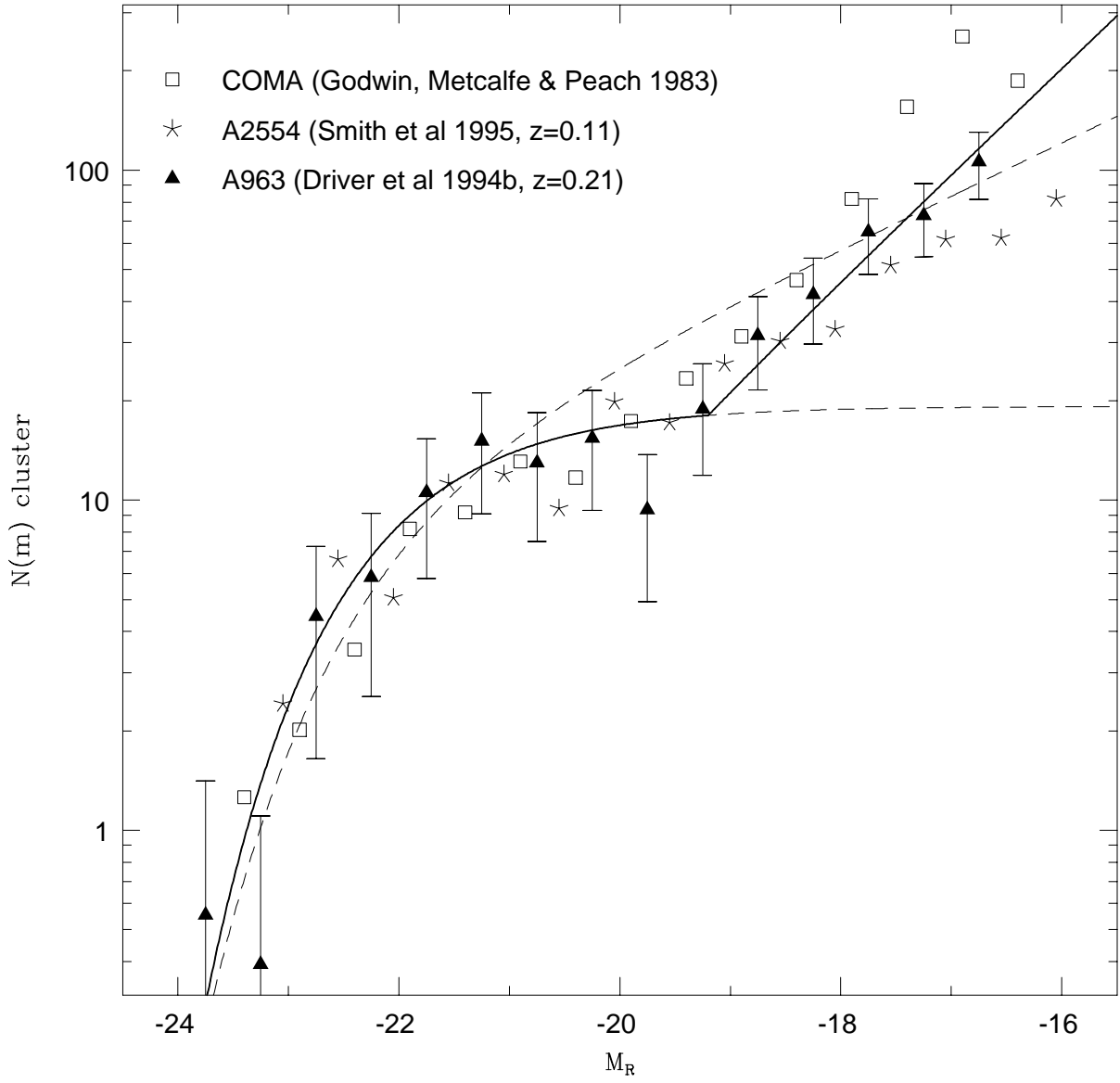

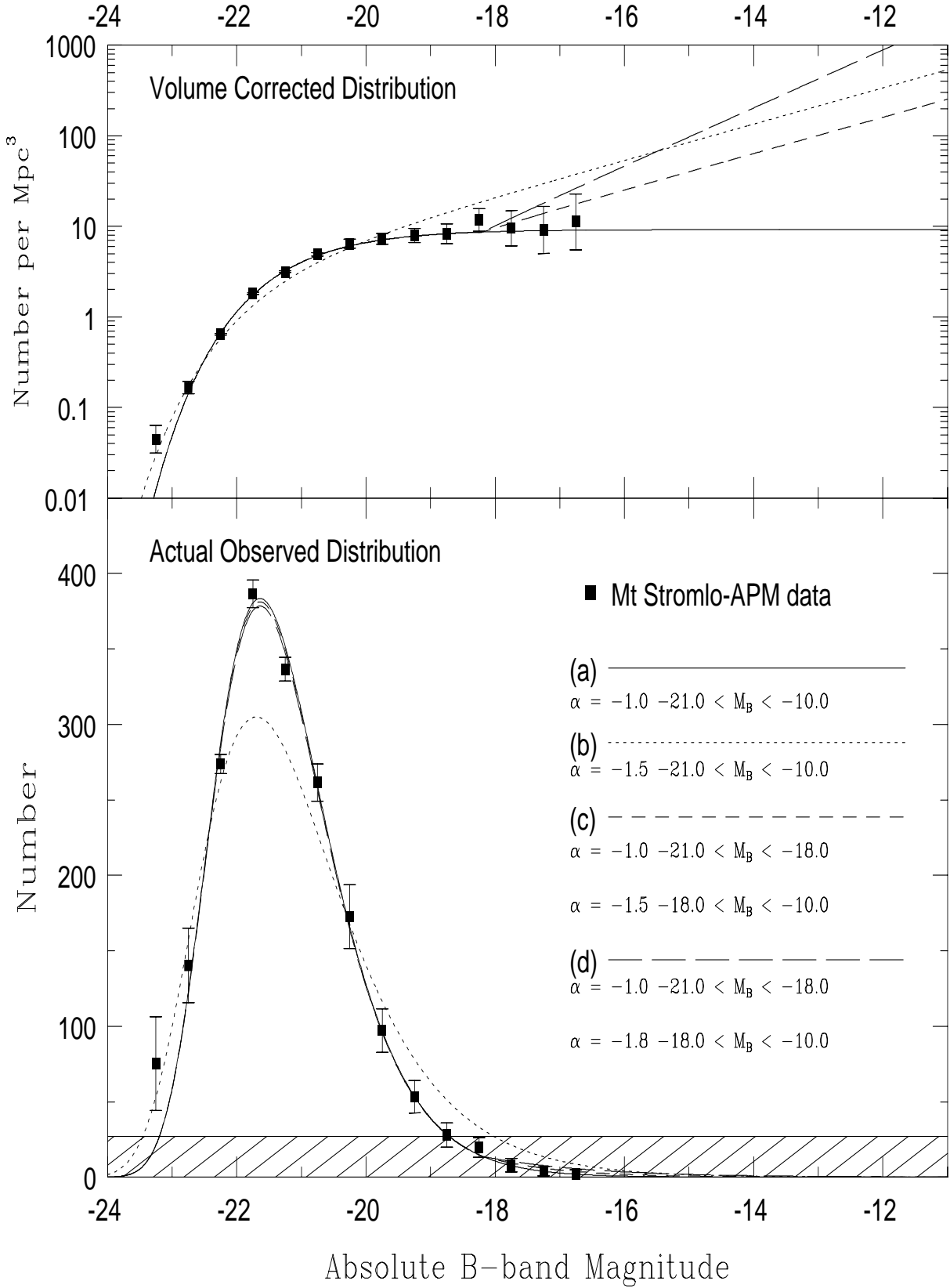

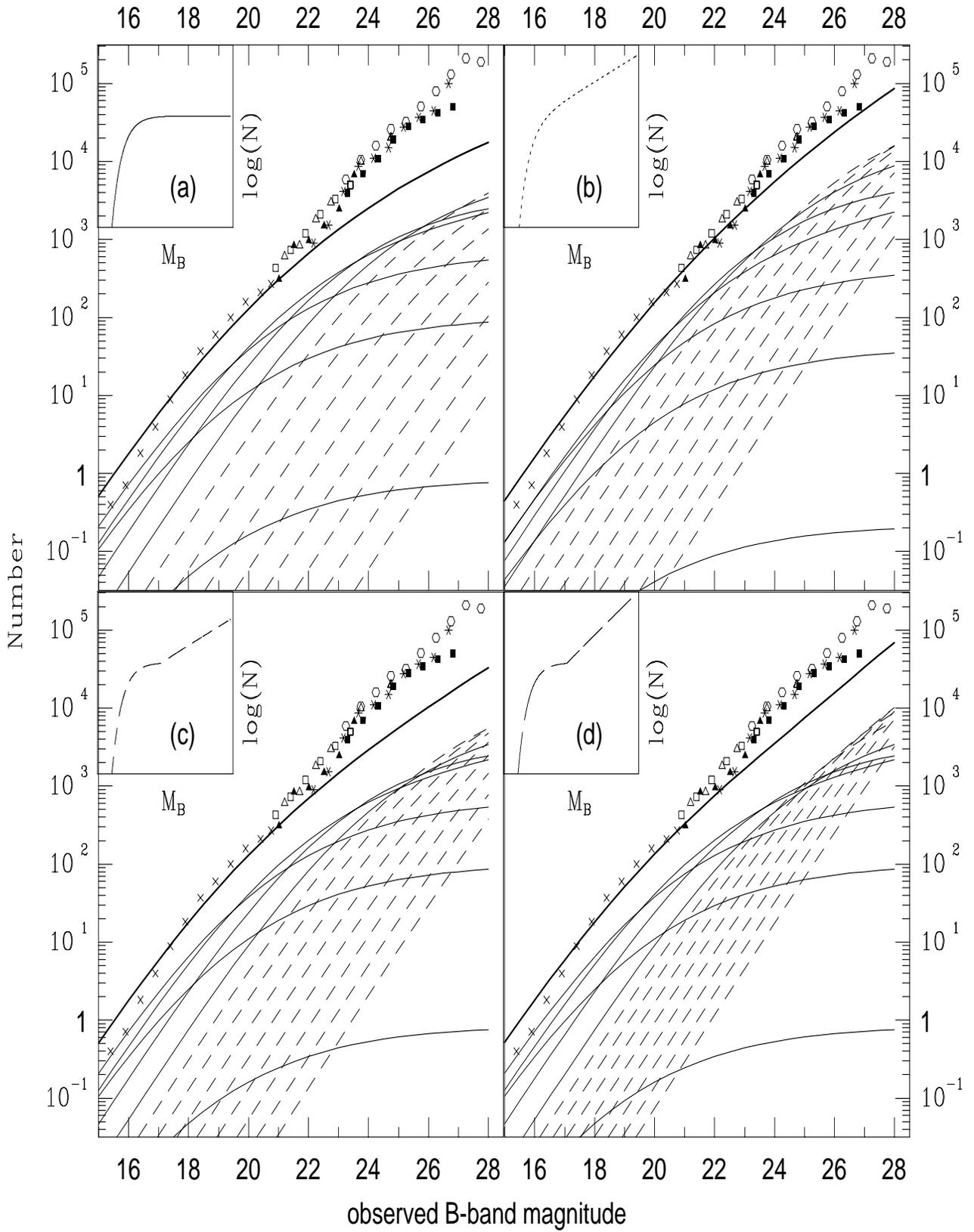

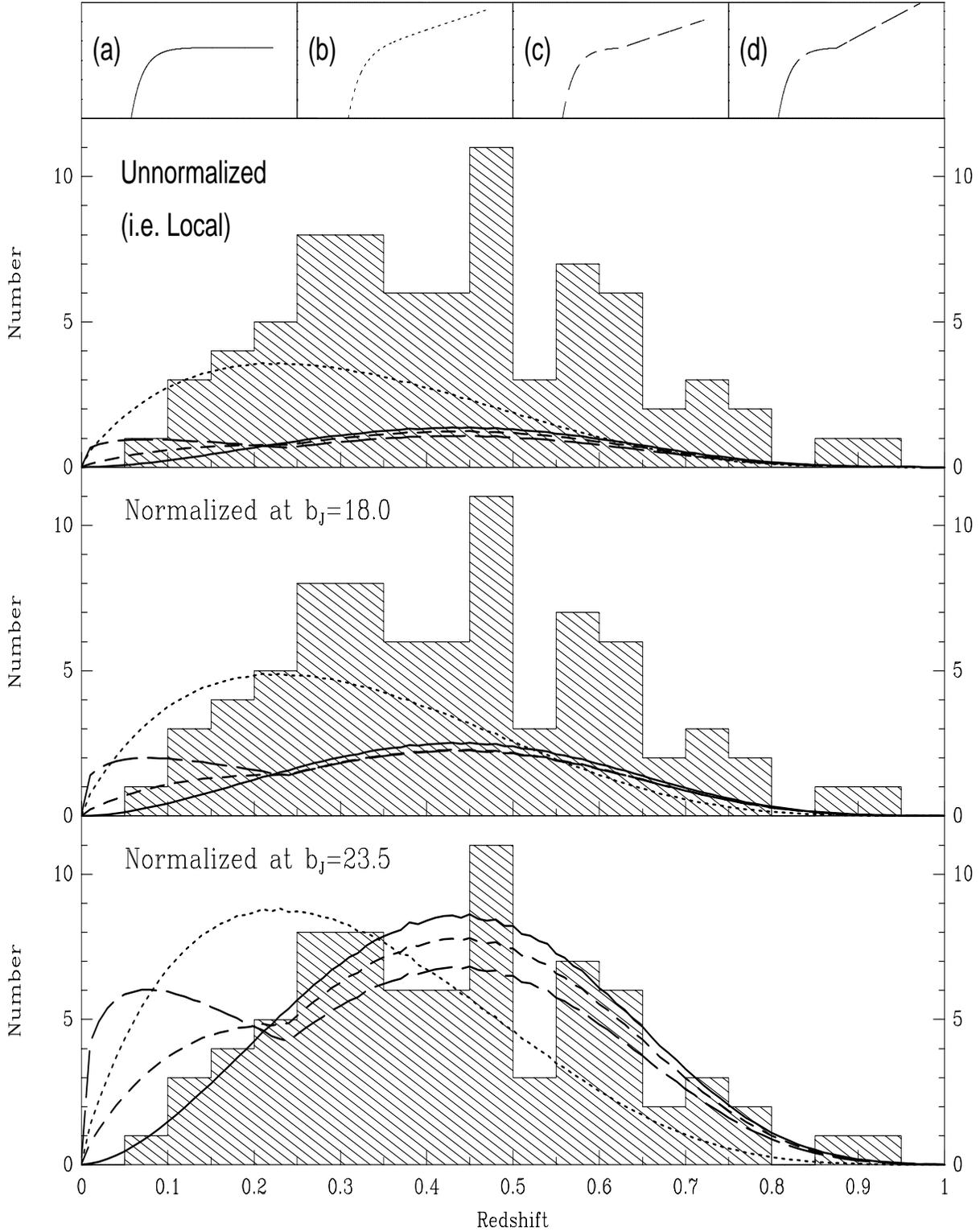